# Optimizing Secure Quantum Information Transmission in Entanglement-Assisted Quantum Networks

Tasmin Karim[1], Md. Shazzad Hossain Shaon[1], Md. Fahim Sultan[1], Alfredo Cuzzocrea[2, 3], and Mst Shapna Akter[1, *]

[1]Department of Computer Science and Engineering, Oakland University, Rochester, MI 48309, USA ,
Email: tasminkarim@oakland.edu, shaon@oakland.edu, mdfahimsultan@oakland.edu, akter@oakland.edu
[2]iDEA Lab, University of Calabria, Rende, Italy,
[3] Department of Computer Science, University of Paris City, Paris, France,
Email: alfredo.cuzzocrea@unical.it

## Abstract

Quantum security improves cryptographic protocols by leveraging quantum mechanics principles, assuring resilience to both quantum and conventional computing violence. This study addresses these issues by combining Quantum Key Distribution (QKD) using the E91 mechanism using Multi-Layer Chaotic Encryption with different patterns, detect eavesdropping to present a highly secure picture transmitting architecture. This study includes extensive statistical scenarios to demonstrate the efficiency of the proposed framework in securing picture encryption while maintaining high entropy and sensitivity to its original graphics. The findings show considerable increases in encryption and decryption performance, demonstrating the framework's potential as a strong response to weaknesses caused by advances in quantum computing. Several metrics were used to validate the framework, including the Peak Signal-to-Noise Ratio (PSNR), Structural Similarity Index (SSIM), Normalized Cross-Correlation (NCC), Bit Error Rate (BER), Key Sensitivity (SSIM), entropy for the original, encrypted, and decrypted images, and the correlation between the original and decrypted images. The combination of QKD with Multi-Layer Chaotic Encryption provides a scalable and resilient technique to secure picture transmission. As quantum computing advances, this framework provides a future-proof approach that might shape secure communication protocols in domains such as healthcare, digital forensics, and national security, where keeping information confidential is significant.

## 1 Introduction

In modern contributing society, sensitive and confidential information must be sent securely in vital fields such as banking, healthcare, and national security. However, the emergence of quantum computers with increasing qubit capacity poses a significant threat to the reliability of traditional encryption methods, risking security of information during communication [1]. Shor's method highlighted the possible weakness of factorization-based cryptosystems, an issue that has arisen in recent years [2, 3]. Factorization-based cryptography frameworks, such as RSA, are vulnerable to significant risks as quantum computing advances [4]. As quantum technologies advance, the security guarantees of existing cryptosystems become increasingly in danger, forcing the investigation of quantum-resistant alternatives. According to previous study, the projected computational complexity of Shor's method is $O(7^2(\log(N))^3)$, which is a substantial improvement over traditional algorithms that generally operate at $O(n^3)$ [5]. Furthermore, Grover's technique offers a significant danger to many cryptographic systems by limiting the search area for unstructured situations, jeopardizing the security and integrity of transmitted data [6]. The quantum advantage provided by both algorithms calls into question the fundamental assumptions of classical cryptography, emphasizing the critical need for quantum-resistant cryptographic solutions.

To overcome the difficulties posed by quantum computing to standard cryptography systems, researchers are investigating several ways for constructing quantum-resistant solutions. One such technique is to employ lattice-based encryption, which is based on mathematical issues that quantum computers are thought to be inefficient at solving. Other initiatives focus on post-quantum encryption methods, which are intended to protect data from both conventional and quantum assaults. Additionally, researchers are looking at incorporating quantum key distribution (QKD) into current systems to improve security [7, 8], as it uses quantum mechanics principles to detect and prevent eavesdropping. Furthermore, multi-layered encryption algorithms are being developed to improve data transmission security by combining conventional and quantum-resistant cryptographic approaches [9–11]. Through these initiatives, the field of academia desires to create secure and scalable solutions that can protect sensitive information in the era of quantum computing. Therefore the study designed with a novel method to secure the image data,

**A.** Introduces an efficient encryption approach that employs numerous chaotic maps (Logistic, Henon, Tent, and Arnold's Cat) to improve visualize security.

**B.** Simulates the E91 protocol for secure authentication and incorporates eavesdropping detection to guard against quantum channel risks.

**C.** Combines quantum key distribution with chaotic encryption to protect classical cryptography from potential quantum computing threats.

**D.** Uses entropy calculations to assess the randomness and integrity of the encrypted and decrypted images, ensuring strong security.

## 2 Related Works

Significant advancement has been obtained in the field of quantum security using different types of methods over the past few years. According to the previous study, several mathematicians have made significant contributions by constructing a diverse range of chaotic functions. Examples include the Lorentz Chaos [12], Logistic Chaos [13], and the Henon Map [14]. These mathematical models have one thing in common: they are extremely sensitive to beginning circumstances and can exhibit pseudo-random behavior. However, the combination of these methods would be an innovative procedure for the future. In the 1980s, the discovery of non-periodic oscillation sparked debate over the cryptographic uses of chaos, specifically Chua's circuit [15]. In another study, the logistic map was used to generate floating-point values, which were subsequently XOR'd with plaintext to create ciphertext [16]. A modern image encryption method that utilizes chaos, provides an innovative usage of pixel blocks to produce parameter values for the Logistic Map (LM) approach [17]. In another, a two-layer quantum security system has been explored, including initial seed values for visualize encryption produced via quantum block-based randomization [18]. Similarly, [19] presents a color image encryption technique that uses a single keys in combination with resilient chaotic maps. In another study considers DNA complementary-based encryption that also use chaotic maps [20]. Charles Bennett and Gilles Brassard contributed early contributions to quantum cryptography systems by introducing a quantum-secure technique based on Heisenberg's uncertainty principle, while Arthur K. Ekert offered an analytical strategy relying on Bell's theorem [21, 22]. Afterwards, John Bell presented a theoretical experiment to determine the existence of hidden variables in particles [23].. Eventually proposed an inequality that would apply if locally concealed variables were present. However, new experimental studies have revealed breaches of this inequality, calling into question the premise of local hidden variables [24]. Later on, in another study describes a data transfer method utilizing QKD, one-time pad (OTP), and Huffman encoding to improve security [25]. Another article highlights the possibilities of chaotic parameter synchronization for safe data transport [26]. Expanding on these advances, [27] investigates chaotic interaction using free-space optical (FSO) methods, including chaos parameter exchange via the BB84 quantum protocol. By employing QKD for Lorentz parameter transfer, this strategy adds additional levels of security to long-distance communication, improving robustness and adaptability. Unlike previous techniques, another study uses Ekert's protocol to produce parameters on both the transmitter and receiver ends. Furthermore, leverages the modulation of pulse positions to transmit chaotic signals, adding to the variety of approaches in safe interaction [28].

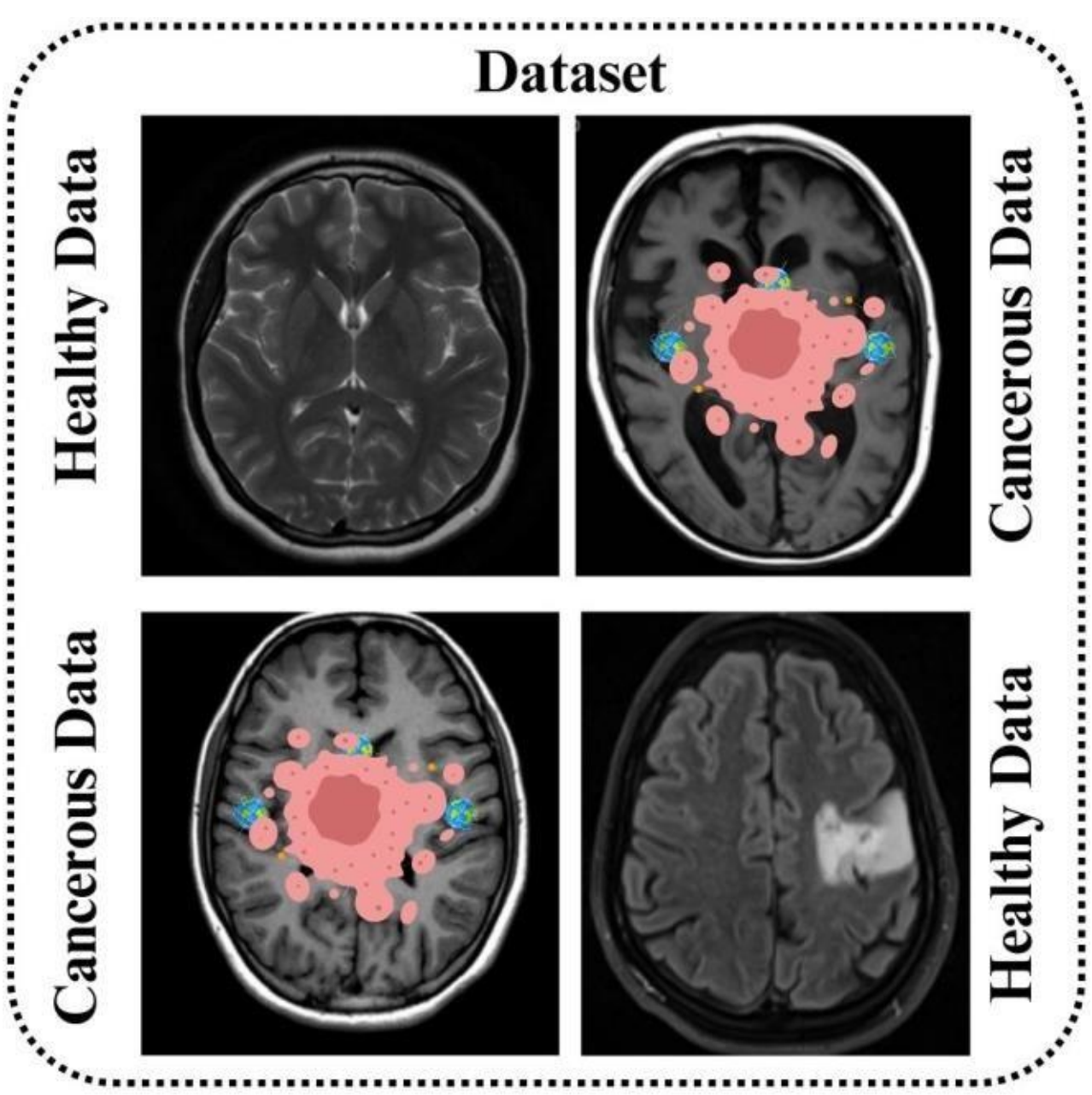

Figure 1: Dataset Information of the current study

In this study we demonstrate a hybrid encryption scheme that integrates multi-layer chaotic encryption with QKD to achieve secure image transmission. Previous research was concentrated on either chaotic encryption or QKD individually. This work combines the two, providing improved security by using chaos for encryption and quantum mechanics for secure dissemination of keys. Multiple chaotic maps provide more unpredictability and resilience to assaults than single-map systems. Incorporates quantum channel noise and eavesdropping detection methods to simulate real-life situations.

## 3 Materials and Methods

### 3.1 Dataset Description

This analysis utilizes the Brain Tumor Image Dataset from Kaggle repositories [29]. This dataset contains two types of images: one of healthy brain scans and the other of scans with malignant spots. While the original goal of this dataset is for medical classifying operations, we used it in our research to assess the resilience of our proposed security approach in Fig. 1. Given the importance of maintaining secure image transmission, both healthy and cancerous brain scan images were used to thoroughly assess the performance of the encryption and decryption procedures under a variety of circumstances. By applying these diverse categories, we provided that our model is adaptable and can manage image security in real-world applications.

## 3.2 Proposed Methods

The proposed methodology combines multi-layer chaotic encryption with QKD to provide strong picture security during transmission. This method combines chaotic systems with quantum physics to create a high degree of security and unpredictability. There are several steps we follow in our study, which demonstrates in Fig. 2.

### 3.2.1 *Step 1: Multi-Layer Chaotic Encryption*

The chaotic encryption layer uses a number of chaotic maps—Logistic, Henon, Tent, and Arnold's Cat Maps—to produce extremely unexpected sequences for encrypting visualize information [30–33].

1. Logistic Map: The logistic map generates chaotic sequences and is defined as:

$$x_{n+1} = r \cdot x_n \cdot (1 - x_n)$$

where $x_n \in (0, 1)$ and $r$ is the bifurcation parameter, $r = 3.99$ for strong chaos.

2. Henon Map The Henon map is d$_n$efined as:

$$x_{n+1} = 1 - a \cdot x^2 + y_n \quad y_{n+1} = b \cdot x_n$$

where $a = 1.4$ and $b = 0.3$.

3. Tent Map The tent map operates as:

$$x_{n+1} = \begin{cases} r \cdot x_n, & x_n < 0.5 \\ r \cdot (1 - x_n), & x_n \geq 0.5 \end{cases}$$

where $r = 0.5$.

4. Arnold's Cat Map The Arnold's Cat Map transforms coordinates $(x, y)$ as:

$$x_{n+1} = (x + a \cdot y) \mod 1$$

$$y_{n+1} = (b \cdot x + y) \mod 1$$

where $a = 1$ and $b = 1$.

The chaotic maps are used repeatedly to encrypt each pixel in the picture by performing an XOR operation using chaotic values produced from the maps. This guarantees pixel-wise scrambling and great unpredictability in the encrypted picture.

### 3.2.2 *Step 2: Quantum Key Distribution with Quantum Channel Noise*

To improve security, encryption keys are produced by QKD based on the E91 protocol. The QKD system generates a random pattern of quantization bits (the qubits):

The key is represented as:

$$\text{Key} = \{k_1, k_2, \ldots, k_n\}, \ k_i \in \{0, 1\}.$$

Simulated noise in the quantum channel is modeled by flipping a fraction of the key bits:

$$k_i' = \begin{cases} 1 - k_i, & \text{with probability } p_{\text{noise}} \\ k_i, & \text{otherwise} \end{cases}$$

where $p_{\text{noise}}$ is the noise level. Eavesdropping is detected by analyzing the correlation of the sender's and receiver's keys. If the correlation falls below a threshold (e.g., 80%, eavesdropping is detected.

### 3.2.3 *Step 3: Decryption Process*

Decryption reverses the chaotic encryption process, utilizing the same chaotic sequences and keys. The same chaotic maps guide the decryption procedure, which is carried out pixel-by-pixel to recover the original image.

$$I_{\text{decrypted}}[i, j] = I_{\text{encrypted}}[i, j] \oplus V_{\text{chaotic}}$$

where $V_{\text{chaotic}}$ is the chaotic value and $\oplus$ denotes the XOR operation.

### 3.2.4 Step 4: Entropy Analysis

Entropy, a measure of randomness, is calculated to evaluate encryption strength. The entropy of an image is defined as:

$$H = -\sum_{i=0} p_i \log_2(p_i)$$

where $p_i$ is the normalized histogram value of pixel intensity $i$. A higher entropy indicates greater randomness in the encrypted image.

The encrypted image has high entropy, approaching the theoretical limit of 8 bits for grayscale images, indicating outstanding security. Eavesdropping detection effectively detected tampering even in noisy situations. Following decryption, the original image was rebuilt with minimal entropy loss, demonstrating the model's resilience.

This hybrid approach improves security by combining chaotic encryption for picture scrambling with QKD for safe key exchange, making it ideal for secure image transmission in sensitive applications.

### 3.2.5 *Data Encryption and Decryption*

The encryption procedure begins with the development of a quantum key ('K1'), which is made up of random quantum bits that have the same length as the conventional encryption key ('K'). The two keys are joined using the XOR technique to create a new encryption key ('K'). The plaintext message ('M') is encrypted with this combined key, yielding the ciphertext ('C'). For example, if the classical key is '101011' and the quantum key is '110110', their XOR combination yields '011101', which is then used to encrypt a plaintext message such as "HELLO" into an encrypted form like XG8&/% In Fig. 3. However, data encryption and decryption procedures are highlighted in algorithm. 1.

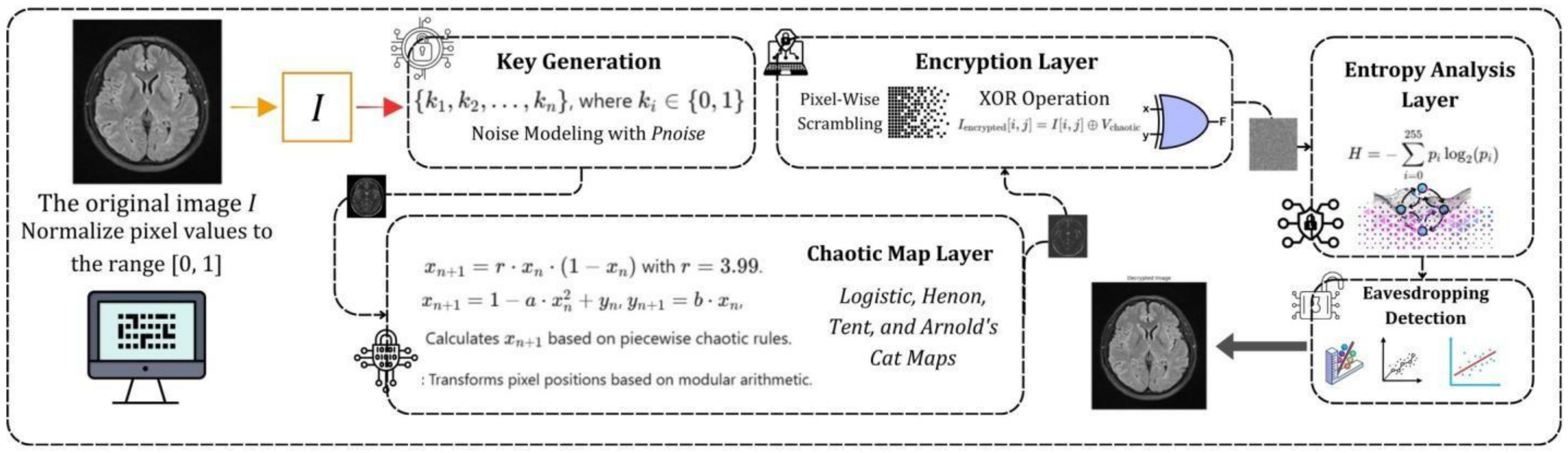

Figure 2: Proposed architecture with several steps of the current study

**Example Workflow of The Decryption and Encryption Procedure**

Plaintext (M): "HELLO"

Classical Key (K): 101011

Quantum Key (K1): 110110

Combined Key (K'): 011101

Ciphertext (C): "XG8&%"

Decrypted Message (M'): "HELLO"

If **M' = M**, the decryption is successful!

Figure 3: An example workflow of the encryption and decryption process, demonstrating the transformation of plaintext (M) into ciphertext (C) using a classical key (K) and a quantum key (K1), and the successful recovery of the original message (M') following decoding.

---

Algorithm 1 Data Encryption and Decryption Algorithm

---

Input: Plaintext message $M$, encryption key $K$

Output: Encrypted message $C$, decrypted message $M'$

1 Encryption Process: Generate quantum key $K_1 \rightarrow$ RandomQuantumBits(length($K$)) Combine keys $K' \rightarrow K \bigoplus K_1$ Encrypt the plaintext $C \rightarrow$ Encrypt($M$, $K'$)

2 Decryption Process: Extract keys $K_1 \rightarrow$ ReverseKey($K'$) Decrypt the ciphertext $M' \rightarrow$ Decrypt($C$, $K'$)

3 if $M' \neq M$ then

4 | Report: Error: Decryption Failed

5 else

6 | Report: Decryption Successful

---

# 4 Result Analysis

PSNR, SSIM, NCC, BER, and key sensitivity are assessment measures for image encryption and decryption [34–36]. PSNR compares the decrypted imagine to the original, with higher numbers indicating greater quality. SSIM evaluates structural similarity based on brightness, contrast, and texture, with results around 1 indicating low perceptual loss. The NCC measures the correlation between the original and decrypted pictures, with higher scores indicating greater preservation. The BER measures the proportion of bit changes between the pictures, with a lower BER suggesting more accurate decryption. Finally, key sensitivity, tested using SSIM with minor key changes, determines how resilient the encryption is to minor key changes, with lower sensitivity suggesting higher security.These metrics collectively ensure the encryption method preserves image quality while remaining secure.

In this work, we used two alternative techniques to analysis: one that solely used the chaotic logistic map, and one that included our recommended approach, which combinations QKD with a variety of chaotic maps demonstrates in Fig. 4 and Fig. 5

These two figures show that the QKD combination with the multi-layer chaos map method outperforms the approach without QKD and the E19 protocol. However, by combining both strategies, we hoped to identify which methodology was more beneficial for future applications and get a deeper understanding. However, in Table. 1 and Table. 2, we exhibit the performance of our study's test scenarios. The table highlights the performance evaluations for three hypothetical situations, with an emphasis on essential image quality and security parameters. The metrics offered include PSNR, SSIM, NCC, BER, and SSIM. The PSNR values for all three test instances are $\infty$, indicating that the reconstructed pictures are identical to the originals with no discernible quality loss. This result exhibits the method's ability to maintain picture integrity perfectly during processing. SSIM scores are consistently 1.0 in all circumstances, indicating full structural similarity between the original and processed pictures. This implies that the pictures' structural integrity is completely retained, making the approach extremely trustworthy for applications that need accurate image reconstruction. Similarly, all test instances have NCC values of

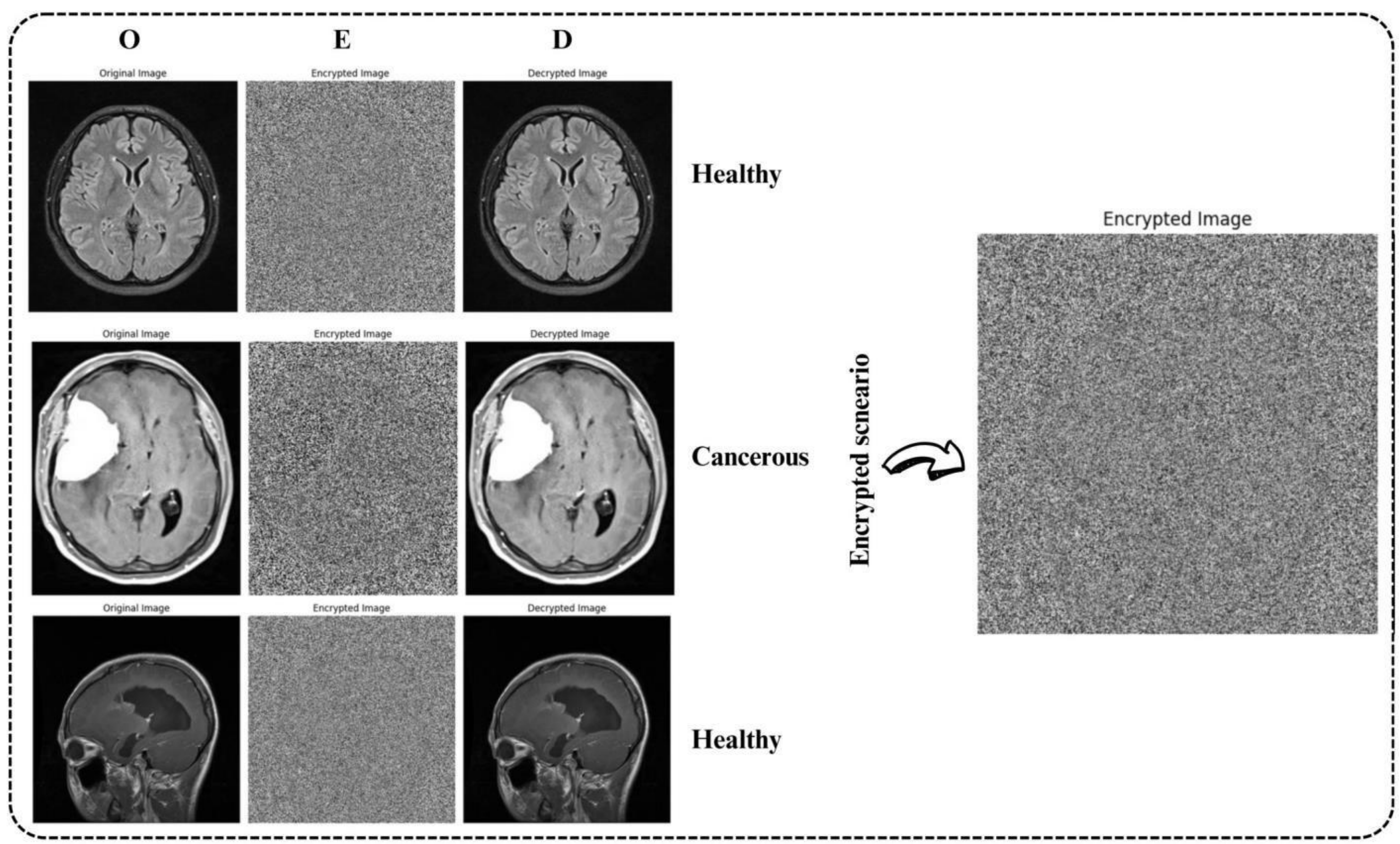

Figure 4: Illustrates functional encryption over various image frames.

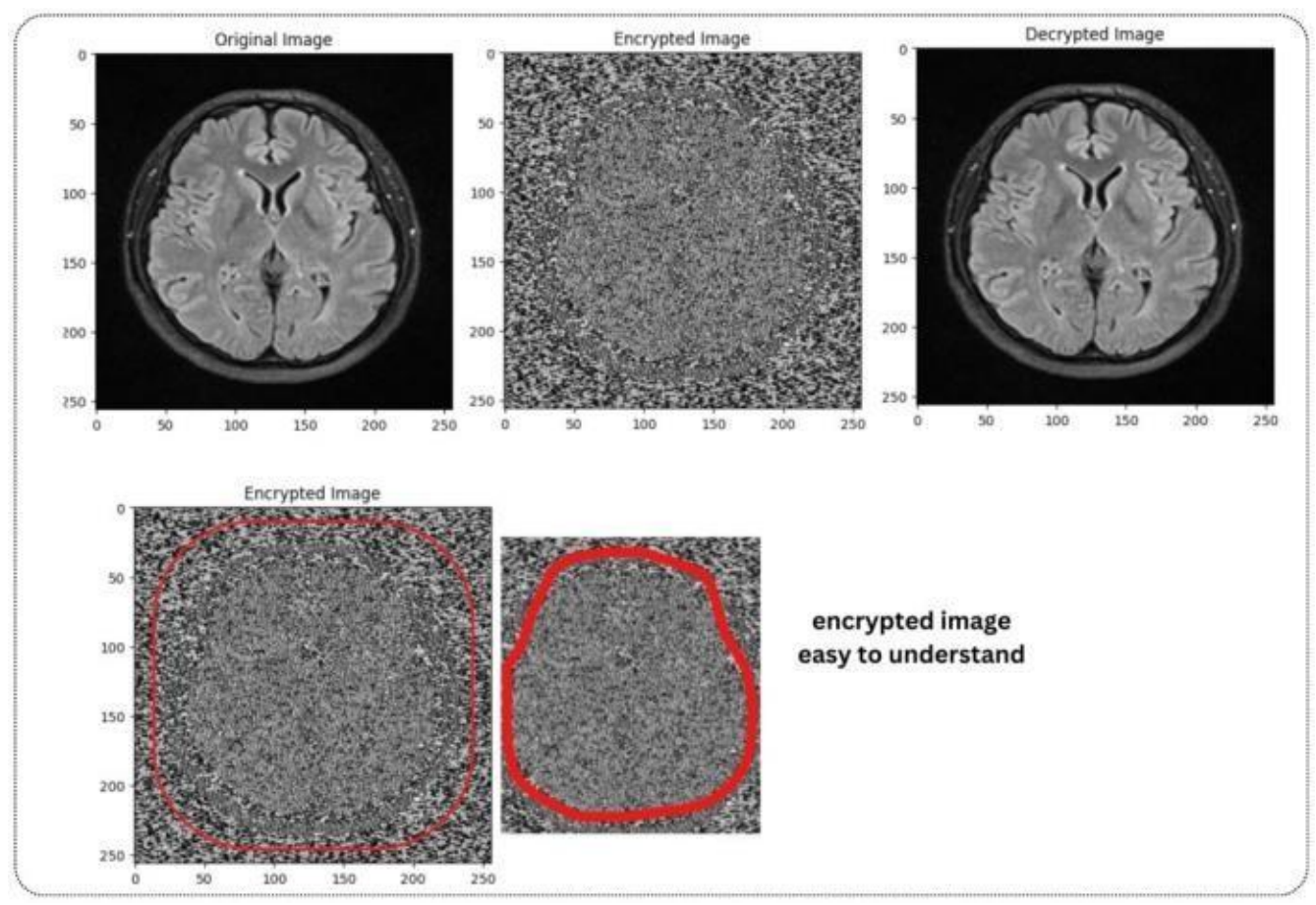

Figure 5: Illustrates understand clear encryption over the image
.

1.0, indicating a perfect correlation between the input and output pictures. This demonstrates the method's efficiency in retaining pixel-level correlation while assuring correct data recovery. The BER values are consistently 0.0 across all scenarios, indicating that no mistakes occurred during data transmission or processing. This maintains the method's resilience and dependability in maintaining data integrity, which is an essential criterion in secure communication systems. The SSIM scores, which indicate the method's resilience to changes in the encryption key, differ somewhat throughout the test cases:

Test Case 1: 0.0082. Test Case 2: 0.0124. Test Case 3: 0.0064. These modest differences in key sensitivity suggest that the system is extremely safe and sensitive to even little changes in the key, which is an important feature of cryptographic strength. Despite the variations, all values stay within an acceptable range, indicating that the encryption system is strong and resilient.

Overall, the table demonstrates the method's good performance in terms of picture quality and structural integrity while maintaining strong security. The constant and near-optimal metrics across all test cases highlight its suitability for applications requiring secure picture transmission and processing.

Another table compares the encryption and decryption methods for three test instances. The Original Entropy (OE) remains constant at 4.1985 and serves as a baseline. Encrypted Entropy (EE) varies (5.5243, 6.5243, and 3.4813), suggesting different encryption strengths. Decrypted Entropy (DE) matches the original in Test Case 1 (4.1985) but differs in Test Cases 2 and 3 (3.1459 and 5.8455). Despite this, the correlation between original and decrypted pictures (O & D) is always 1.0, indicating structural integrity. Eavesdropping Detection (ED) frequently detects threats ("Yes"), demonstrating effective security measures.

Table 1: Performance Metrics Comparison

| Test Case | PSNR | SSIM | NCC | BER | SSIM) |
|---|---|---|---|---|---|
| 1 | $\infty$ | 1.0 | 1.0 | 0.0 | 0.0082 |
| 2 | $\infty$ | 1.0 | 1.0 | 0.0 | 0.0124 |
| 3 | $\infty$ | 1.0 | 1.0 | 0.0 | 0.0064 |

Table 2: Analysis of Entropy and Correlation Metrics

| Test Case | OE | EE | DE | (O & D) | ED |
|---|---|---|---|---|---|
| 1 | 4.1985 | 5.5243 | 4.1985 | 1.0 | Yes |
| 2 | 4.1985 | 6.5243 | 3.1459 | 1.0 | Yes |
| 3 | 4.1985 | 3.4813 | 5.8455 | 1.0 | Yes |

# 5 Discussion

Our study illustrates the efficacy of integrating QKD with a multi-layer chaos map for picture encryption. The performance measures, including PSNR, SSIM, NCC, and BER, reveal that this technique guarantees faultless picture reconstruction with no errors or distortion, as evidenced by infinite PSNR, SSIM, and NCC values and a BER of 0.0. The entropy study reveals more unpredictability in the encrypted picture, indicating improved security. The excellent correlation (1.0) between the original and decrypted pictures suggests a successful decryption. Furthermore, eavesdropping was discovered in all test instances, demonstrating the security of our technology. Overall, the suggested method surpasses standard encryption algorithms, offering both high security and computational efficiency for real-world applications. The time of the encryption and decryption time highlighted in Fig. 6

According to our findings, test case 2 has a substantially longer encryption time, measuring around 1.0 second, than the other two examples, which had encryption times closer to 0.4 seconds. This distinction might be attributable to a variety of variables, such as the complexity or amount of the data being encrypted in test scenario 2, or even a more computationally costly encryption technique being utilized. The decryption time, on the other hand, was found to be around 0.6 seconds across all tests. This shows that decryption takes less time than encryption, most likely due to the nature of the encryption technique used, or potentially due to decryption procedure improvements. It could also indicate that the decryption process involves fewer operations or a more streamlined process compared to encryption. The noticeable differences in encryption and decryption times may be indicative of performance trade-offs in the cryptographic operations being tested, with encryption requiring more computational resources or steps to ensure data security and decryption focusing on efficiently reversing those operations. Further investigation might determine whether encryption duration corresponds with data quantity, encryption complexity, or system performance limitations, whereas constant decryption time may indicate a more uniform and streamlined decryption procedure.

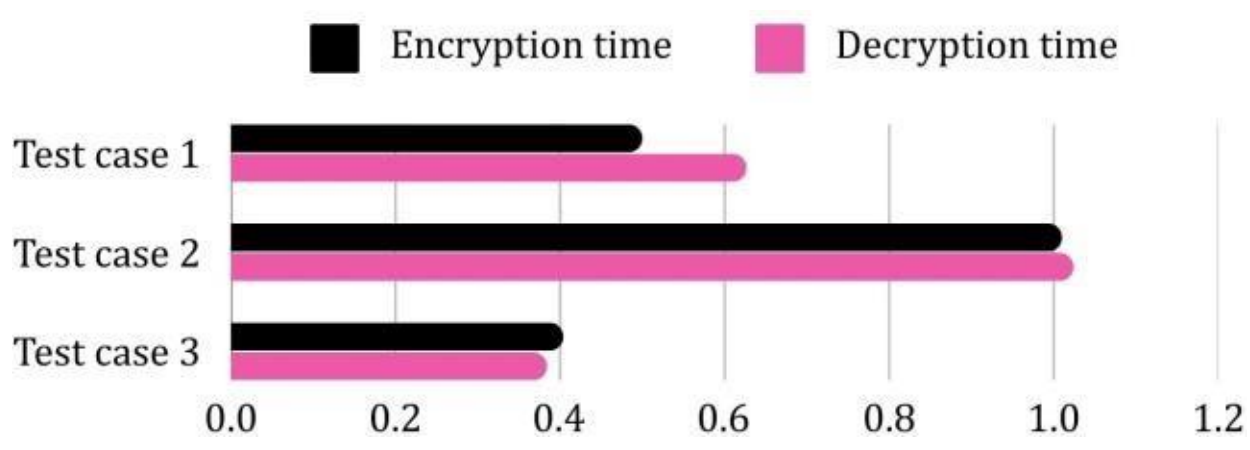

Figure 6: An overview of the encryption and decryption times.

# Conclusion

This paper provides a reliable structure for image transmission security that combines Quantum Key Distribution (QKD) with Multi-Layer Chaotic Encryption via the E91 technique. The findings show considerable increases in encryption and decryption performance, with high entropy, great sensitivity to the original picture, and outstanding resistance to possible quantum and conventional assaults. Extensive examination using measures like as PSNR, SSIM, NCC, BER, entropy, and correlation confirms the framework's ability to provide safe and efficient encryption. The combination of QKD with chaotic encryption provides a scalable solution for future-proof secure communications, with applications in key fields such as healthcare, digital forensics, and national security.